\documentclass[11pt,twoside]{article}
\usepackage{FUSE2004}
\usepackage{natbib}

\usepackage{epsf}
\usepackage{psfig}
\usepackage{lscape}

\begin{document}
\title{FUSE and the Astrophysics of AGN and QSOs}
\author{Gerard A. Kriss}
\affil{Space Telescope Science Institute, 3700 San Martin Drive,
Baltimore, MD 21218, USA\\email: gak@stsci.edu}

\begin{abstract}
The high spectral resolution and sensitivity of {\it FUSE} have enabled 
far-ultraviolet studies of AGN and QSOs that are a natural complement to 
observations using {\it HST, Chandra,} and {\it XMM-Newton}.
Through synergistic use of the large sample of nearby AGN that
serve as background probes of gas in the Galactic halo and the ISM, the
{\it FUSE} PI team has observed a large number (approaching 100) of the nearest
and brightest AGN. In addition to emission from {\sc O~vi}, we identify emission
lines due to {\sc C~iii}, {\sc N~iii}, {\sc S~iv}, and He {\sc ii}
in many of the Type-1 AGN.
More than half of the Type 1 objects also show intrinsic absorption by the
{\sc O~vi} doublet as well as {\sc C~iv} absorption and evidence of a soft X-ray
warm absorber.  Guest investigators have 
successfully coordinated {\it FUSE} observations of bright AGN with 
simultaneous {\it HST} and X-ray observations.
These have contributed greatly to our understanding of the UV and X-ray
absoring gas in AGN as either a wind from the accretion disk,
or a thermally driven wind from the obscuring torus.
\end{abstract}

\section{Introduction}

Observations of the nearest and brightest active galaxies have inspired our
current paradigm for the workings of active galactic nuclei (AGN).
Their proximity gives superb spatial scale, which has been exploited
in imaging with the {\it Hubble Space Telescope (HST)}
and in high-resolution radio observations.
These same nearby AGN have the highest S/N {\it HST} far-UV spectra,
and the best high-resolution X-ray spectra.
The spectral energy distribution of AGN peaks in
the far-ultraviolet wavelength range.
Thus, it is important for understanding the energy generation mechanism
and the processes that govern accretion onto massive black holes.
Since this portion of the spectrum also determines the radiative input to the
broad-line region (BLR) and the narrow-line region (NLR) in AGN,
as well as the surrounding host galaxy and the intergalactic medium (IGM),
determining its spectral shape
is a crucial input for understanding the physical conditions of the gas
in surrounding regions.
The high spectral resolution and high sensitivity of the
{\it Far Ultraviolet Spectroscopic Explorer} ({\it FUSE}) enables us to study
the astrophysics of the nearest AGN in this crucial waveband.

The 900--1200 \AA\ spectral range contains numerous diagnostic spectral
features that can be applied to AGN physics.
The most prominent emission line is the {\sc O~vi} resonance doublet
$\lambda\lambda 1032,1038$, which is particularly
strong in the spectra of low-redshift, low-luminosity AGN
due to the Baldwin effect \citep{Scott04}.
This line can serve as a diagnostic of the energy input from the extreme
ultraviolet to soft X-ray portions of the ionizing continuum.
Likewise, the {\sc O~vi} doublet is a crucial diagnostic for the
absorbing gas commonly seen as {\sc O~vii} and {\sc O~viii} absorption
in X-ray spectra of AGN (\citealt{Reynolds97, George98}), and as
Ly$\alpha$ and {\sc C~iv} absorption in {\it HST} spectra \citep{Crenshaw99}.
The high-order Lyman lines and the Lyman limit provide additional
diagnostics of absorbing gas.
In some cases (e.g., NGC 4151 or NGC 3516) the neutral hydrogen can be
optically thick and thereby play a significant role in collimating the
ionizing radiation that illuminates the NLR \citep{Kriss97}.
Finally, numerous ground-state transitions of molecular hydrogen
in the Lyman and Werner bands provide a sensitive tracer of molecular gas.
Under the right circumstances, one might expect to see $\rm H_2$ associated
with the obscuring torus in AGN in absorption against the continuum and
broad emission lines.

\begin{figure}[t]
\plotfiddle{"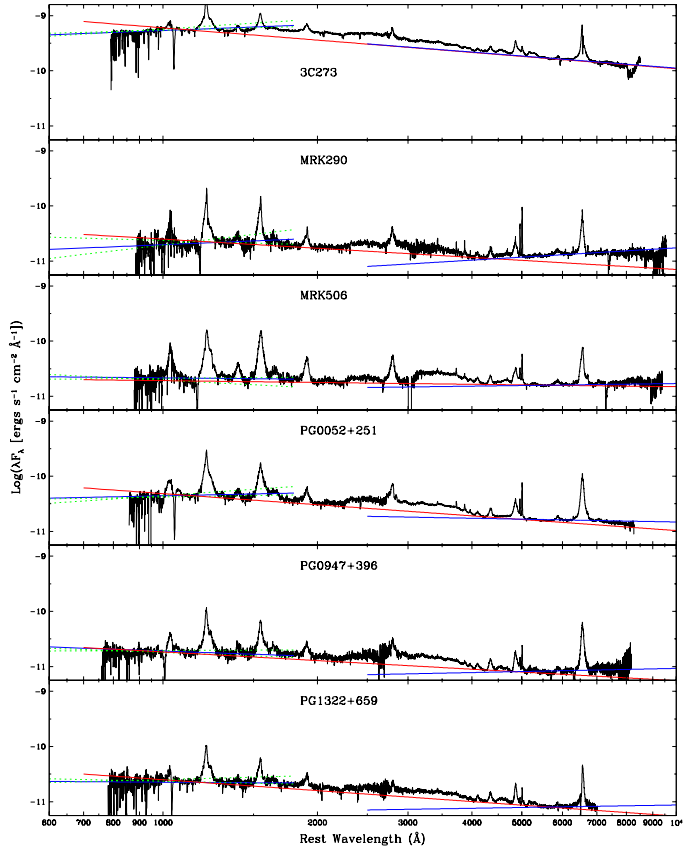"}{4.35in}{0}{45}{45}{-145}{-15}
\caption{
FUV to optical SEDs and fitted power-laws for selected {\it FUSE} AGN.
Dotted lines show the uncertainties in the FUV power-law fits.
\label{uvosedfig}}
\end{figure}

\section{The Ionizing Continuum Shape}

The ionizing continuum shape in AGN is difficult to study directly.
At low redshift it is mostly obscured by our own interstellar medium, and
inferences must be made from the spectral shape observed in the far-UV and
soft X-ray (on either side of the opaque interstellar absorption) as well
as from clues provided by the emission lines observed.
Composite spectra assembled from {\it HST} observations of moderate-redshift quasars
(\citealt{Zheng97, Telfer02}) show that the spectral energy
distribution of AGN peaks in the far-ultraviolet, with a distinct break at
$\sim1000$ \AA.
The short wavelength response of {\it FUSE} allows us to investigate the continuum
properties of lower-redshift, lower-luminosity AGN. We have used both the
composite spectrum approach, and the detailed assembly of spectral energy
distributions for 17 individual AGN.

\begin{figure}[t]
\plotfiddle{"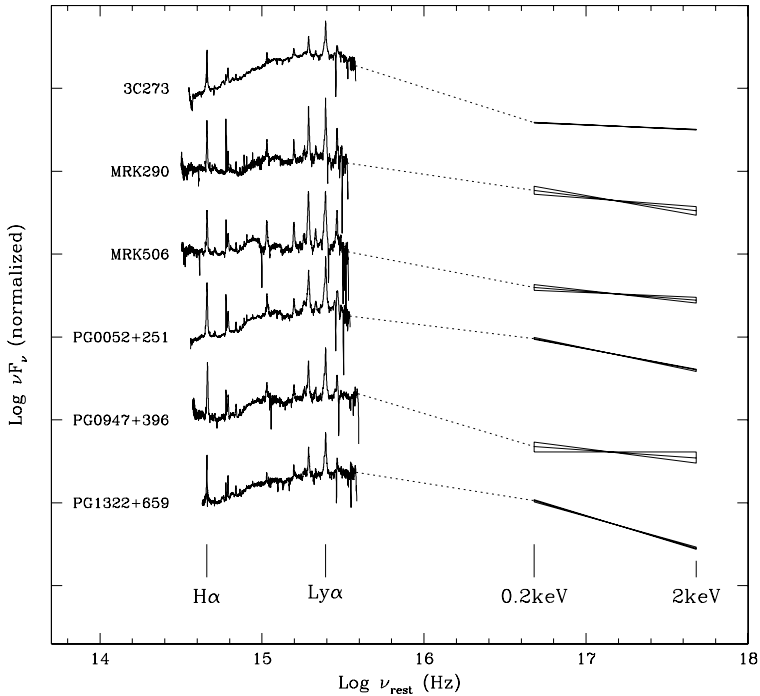"}{3.65in}{0}{50}{50}{-145}{-75}
\caption{
Optical/UV/X-ray spectral energy distributions of representative AGN.
The dotted lines are drawn to connect the FUV and X-ray spectra.
\label{xsedfig}}
\end{figure}

\cite{Shang04} obtained quasi-simultaneous observations of 17 AGN using
{\it FUSE}, {\it HST}, and ground-based telescopes to produce spectral energy
distributions covering 912--9000 \AA\ in the observed frame.
A sampling of these spectra in Figure \ref{uvosedfig} shows that single power
laws adequately describe the UV-optical continuum shape of roughly half the
sample.  Many of the objects, however, show breaks in the spectral index at
$\sim1100$ \AA, similar to the {\it HST} composites. When we compare the
UV-optical spectra with archival X-ray spectra (Figure \ref{xsedfig}),
one can see that the spectral energy
distribution is clearly peaking in the far-ultraviolet. In most cases, the
objects in the \cite{Shang04} sample have an X-ray spectral slope and
normalization matching the extrapolation of the FUV continuum.

\begin{figure}
\plotfiddle{"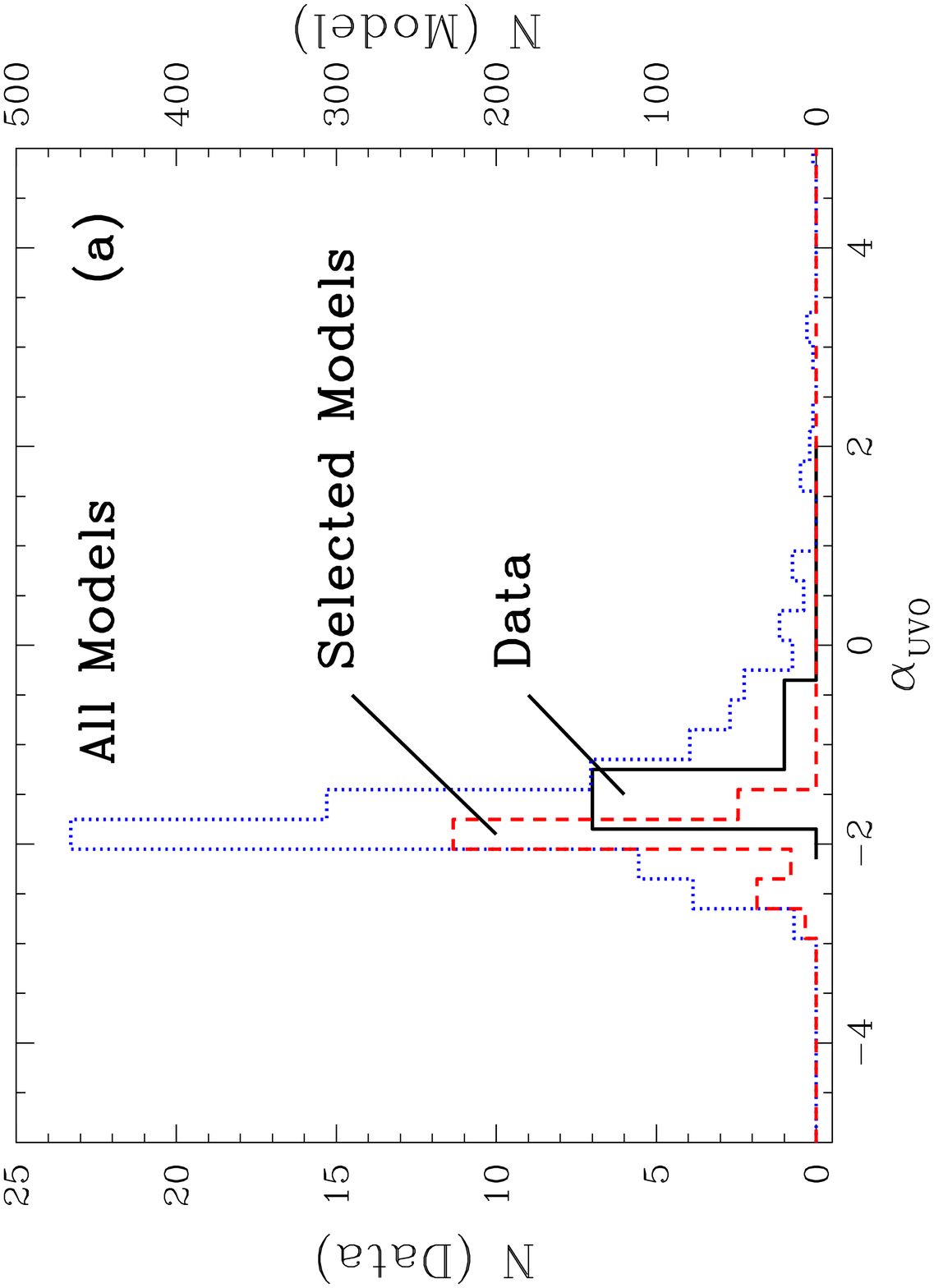"}{2.10in}{-90}{24}{24}{-191}{167}
\vspace{-1.9in}
\plotfiddle{"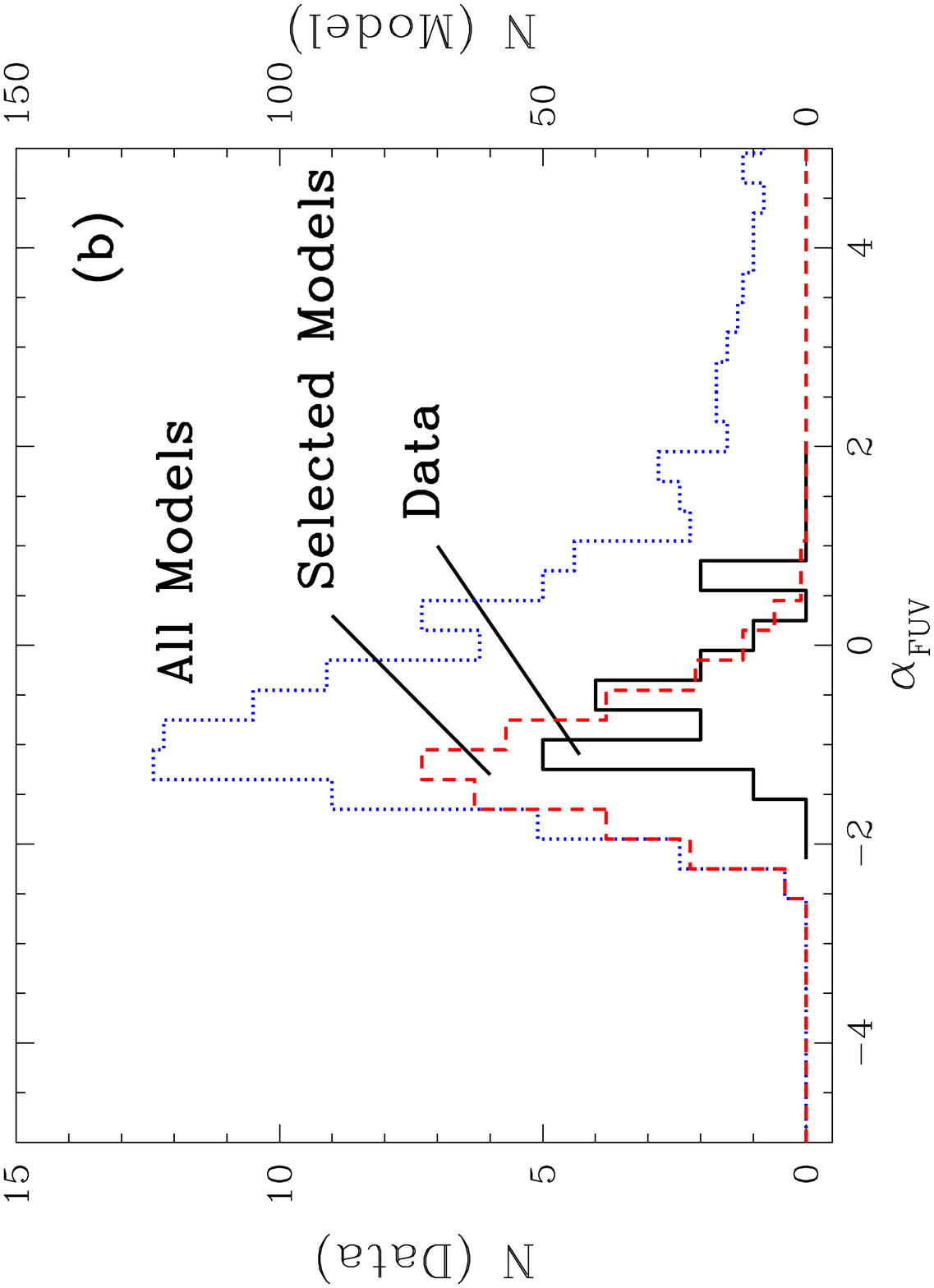"}{2.10in}{-90}{24}{24}{000}{194}
\vspace{-0.8in}
\caption{
Distributions of $\alpha_{uvo}$ and $\alpha_{fuv}$ for the \cite{Shang04}
sample (solid line), all accretion-disk models (dotted line),
and accretion-disk models spanning the same range in mass and Eddington ratio
as the sample.
\label{admodsfig}}
\end{figure}

These individual SEDs are broadly consistent with modern accretion-disk models.
Geometrically thin disk models with non-LTE atmospheres
\citep{Hubeny00} predict a narrow distribution of power-law indices for
disk spectra on the red side of the peak in the SED.
These models peak in the far-ultraviolet, with a spectral break or bump near
the Lyman limit. Spectral indices shortward of the break have a broader
distribution. As shown in Figure \ref{admodsfig} the 17 AGN from \cite{Shang04}
have roughly similar distributions of spectral indices compared to the models.

\begin{figure}[h]
\plotfiddle{"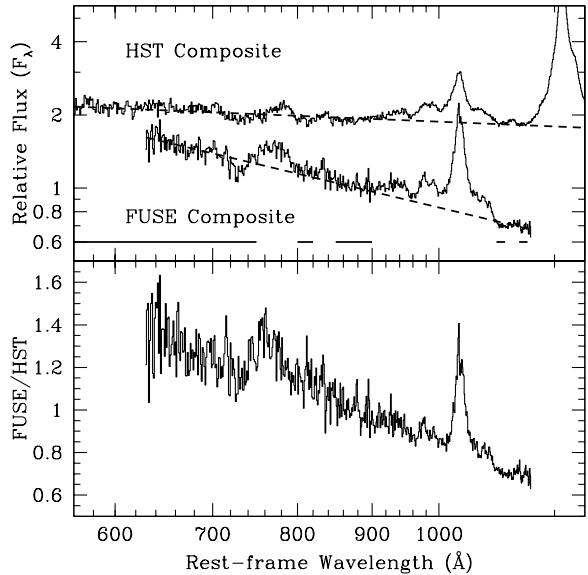"}{2.85in}{0}{40}{40}{-125}{-60}
\caption{
{\it Top panel:} Composite AGN spectra with power law continuum fits
(dashed lines) and wavelength regions used in fit (solid lines).
{\it Bottom panel:} The ratio of the {\it FUSE} composite to the {\it HST}
composite spectrum.
\label{compositefig}}
\end{figure}

With the same technique used to create the {\it HST} composite AGN spectra,
we have constructed a composite {\it FUSE} spectrum from 128 observations of
85 distinct AGN \citep{Scott04}. As shown in Figure \ref{compositefig}, the
{\it FUSE} composite
($\alpha = -0.56^{+0.38}_{-0.28}$ for $f_\nu \sim \nu^{\alpha}$) is
bluer than the {\it HST} composite ($\alpha = -1.76 \pm 0.12$), it shows no
evidence for a break in the spectral index in the far-ultraviolet, and it has
significantly stronger Ne~{\sc viii} and {\sc O~vi} emission.

\begin{figure}
\plottwo{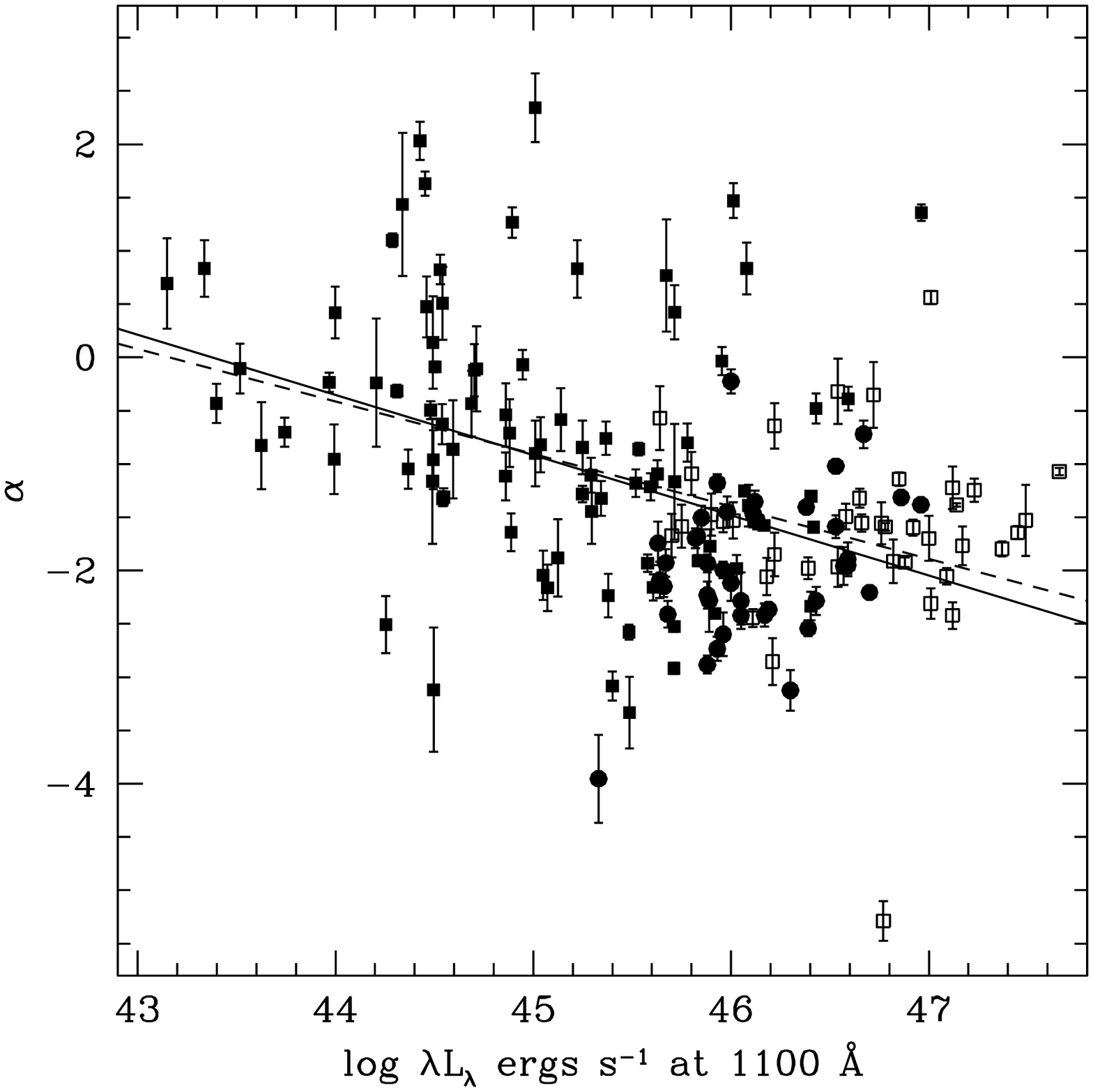}{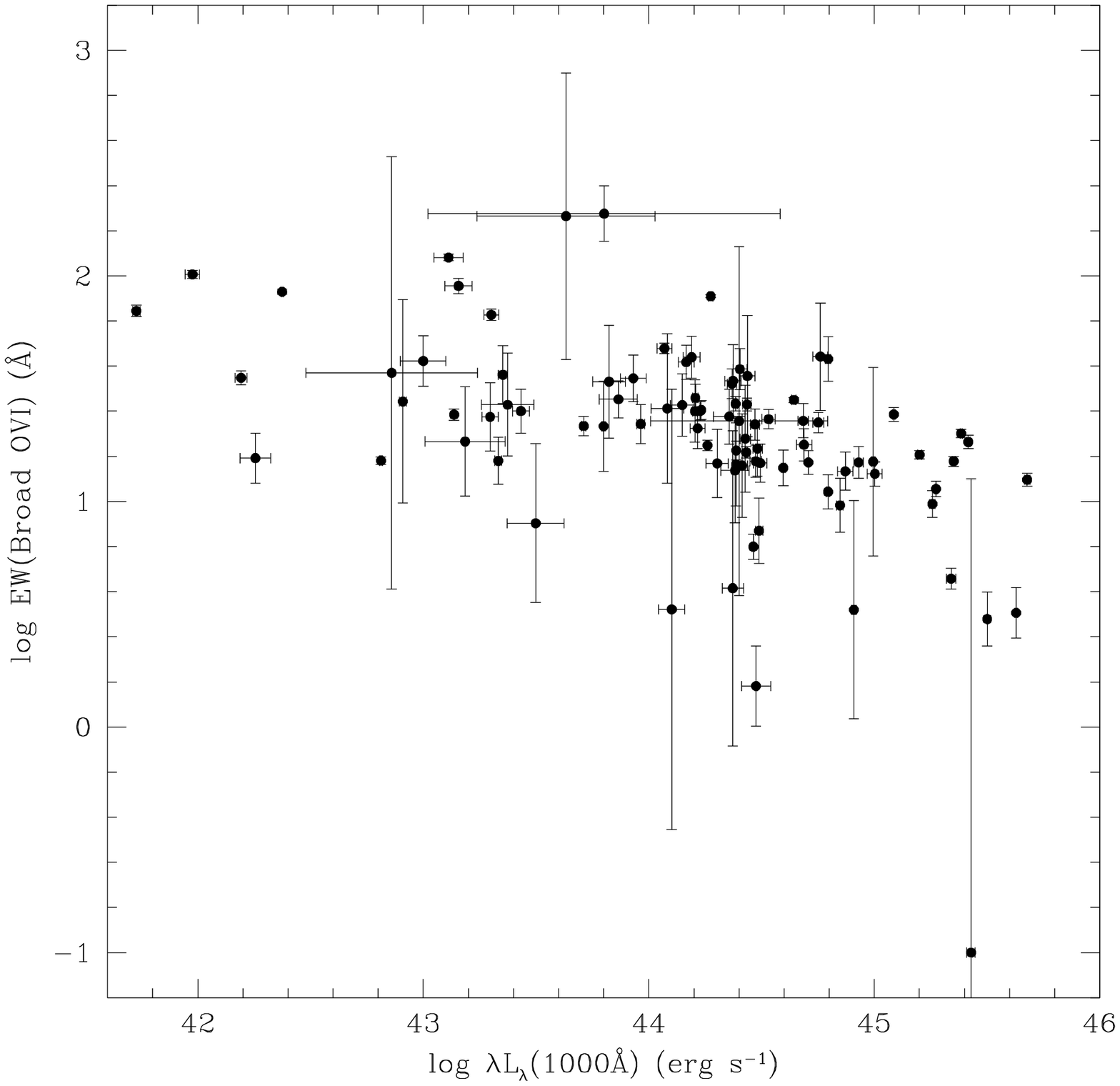}
\caption{
{\it Left Panel:}
EUV spectral index (for $f_{\nu} \sim \nu^{\alpha}$) vs. the 1100 \AA\ luminosity for {\it FUSE} (solid squares) and {\it HST} (Radio-loud: solid circles,
Radio-quiet: open squares) AGN. The dashed line is the best linear fit to the
combined {\it FUSE} and {\it HST} sample; the solid line is for the {\it FUSE}
sample alone.
{\it Right Panel:}
The equivalent width of broad {\sc O~vi} emission in the {\it FUSE} AGN shows a
strong anticorrelation with the continuum luminosity at 1000 \AA.
\label{baldwinfig}}
\end{figure}

In an effort to understand these properties in more detail, we examined the
spectral indices of individual objects in the sample.
As found by \cite{Shang04}, some individual objects do show breaks in
their spectrum, but, in general, the {\it FUSE} AGN are lower luminosity and
bluer than individual {\it HST} AGN. The left panel of Figure \ref{baldwinfig}
shows a significant anticorrelation between the AGN spectral index and
its luminosity.
We interpret this in the context of an accretion disk around the central black
hole in the following manner---lower-luminosity AGN are likely to have less
massive black holes, and hence hotter accretion disks. This shifts the peak of
their spectral energy distributions to shorter wavelengths, moving the spectral
break out of the {\it FUSE} bandpass and making their spectra bluer.
This physical interpretation may also explain the Baldwin effect.
The bluer spectra will have more high-energy photons, leading to relatively more
emission from highly ionized species such as {\sc C~iv}, {\sc O~vi}, and
Ne~{\sc viii}. As shown in the right panel of Figure \ref{baldwinfig},
{\sc O~vi} emission in the {\it FUSE} sample shows a strong anticorrelation
with luminosity.

\section{Emission and Absorption Features in {\it FUSE} AGN Spectra}

As of November 1, 2002, {\it FUSE} spectra for a total of 104 AGN were present
in the archive. Four of these are Type 2 AGN, and the rest are Type 1.
Two of the Type 2 AGN show strong, narrow emission lines (NGC~1068 and Mrk~477),
and the other two have stellar continua typical of starburst galaxies.
53 of the Type 1 AGN have $z < 0.15$, so that the {\sc O~vi} doublet is visible
in the {\it FUSE} band.
As shown earlier in Figure \ref{baldwinfig}, all of these AGN have strong,
broad {\sc O~vi} emission.
Surprisingly, roughly a third of these (17/53) also show strong {\it narrow}
{\sc O~vi} emission. Emission from {\sc C~iii} $\lambda 977$ and
{\sc N~iii} $\lambda 991$ is common, as is a bump of blended emission lines
on the red wing of {\sc O~vi}. In narrow-line Seyfert 1 galaxies such as
I~Zw~1, this bump is resolved into emission from
{\sc S~iv} $\lambda\lambda 1062,1072$ and He~{\sc ii} $\lambda 1085$.

\begin{figure}
\plottwo{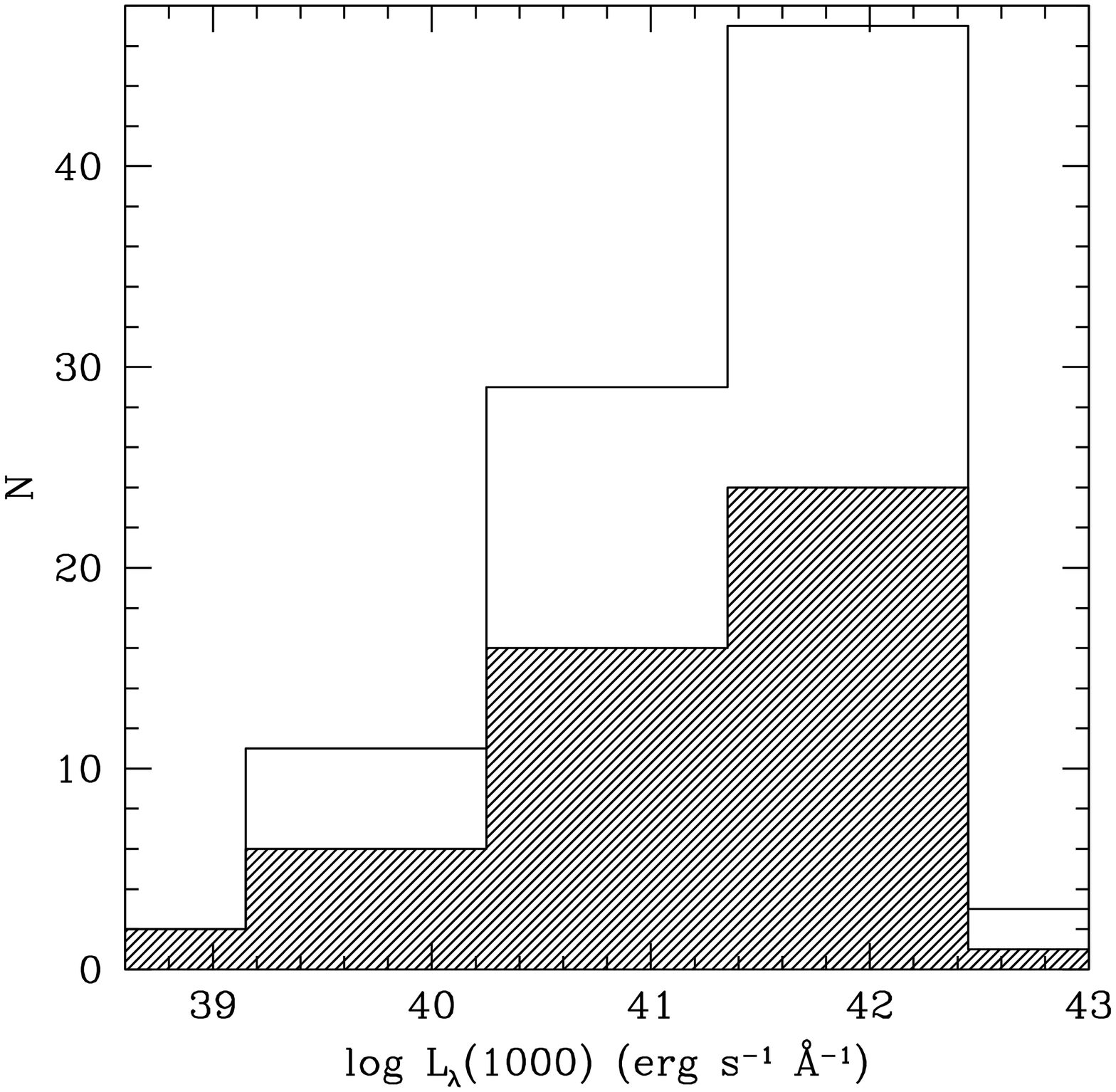}{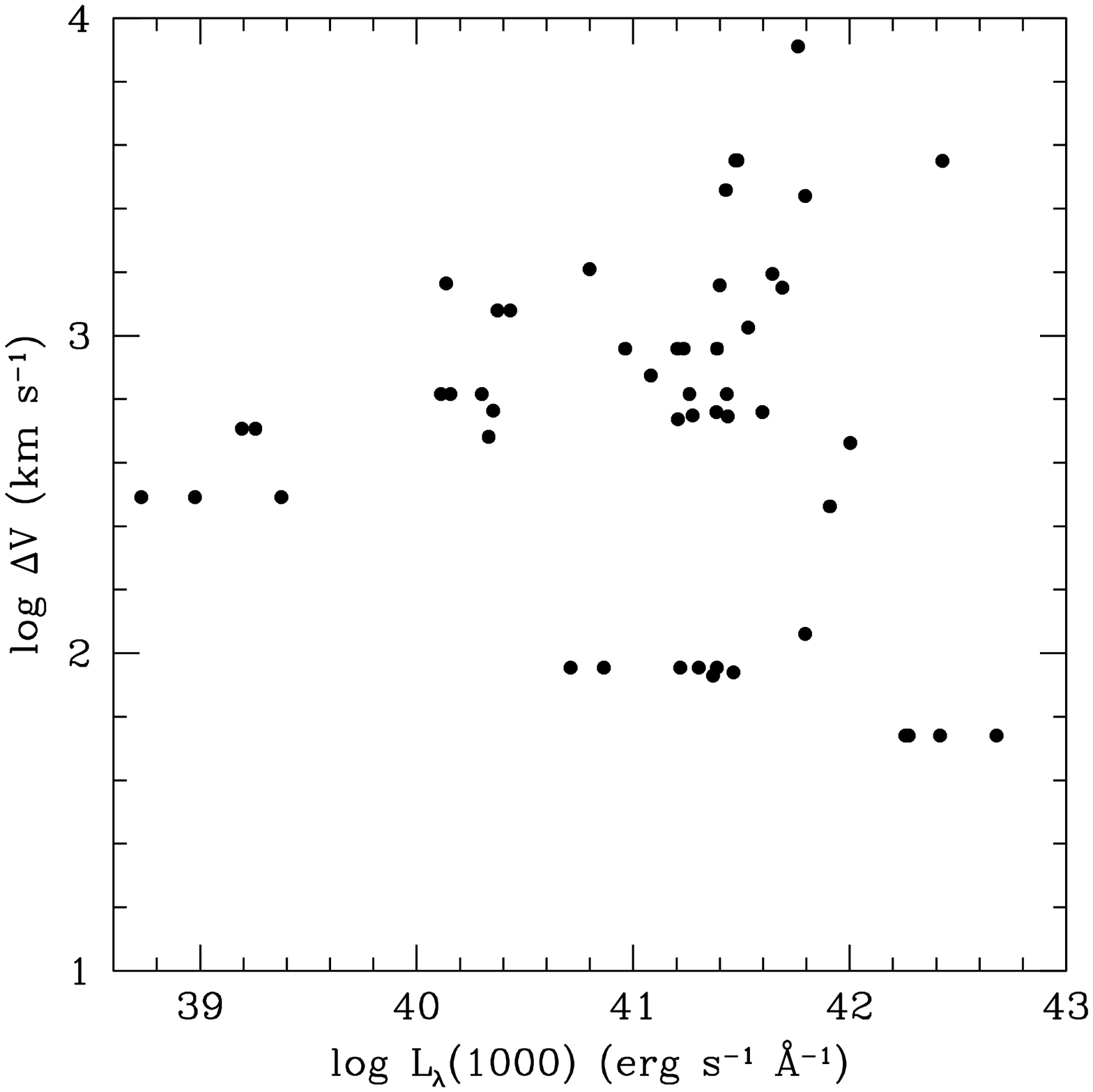}
\caption{
Left: Histogram of {\it FUSE} AGN versus luminosity. The shaded area shows the number 
of objects exhibiting intrinsic absorption. Right: The points show outflow
velocity as a function of luminosity.
\label{abshistfig}}
\end{figure}

\begin{figure}
\plotfiddle{"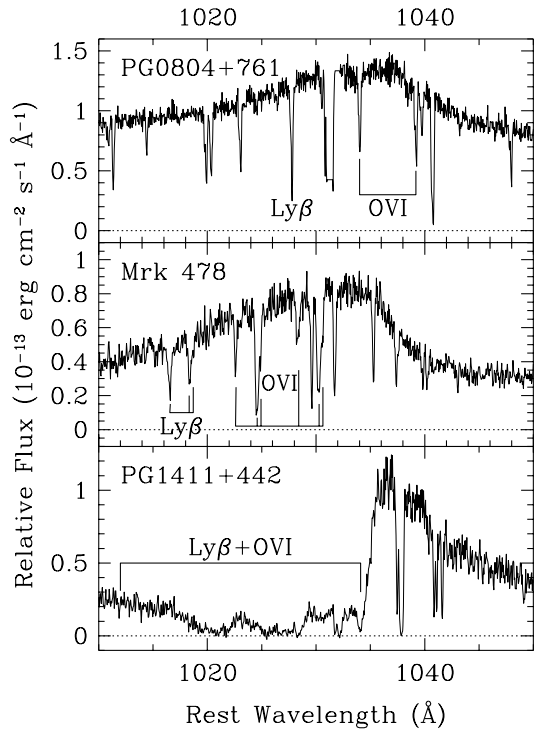"}{3.55in}{0}{40}{40}{-115}{-32}
\caption{
The three classes of {\sc O~vi} absorption line morphology: Single, isolated
lines: PG0804+761 (top); Multiple, blended lines:
Mrk~478 (middle); Broad, blended
trough: PG1411+442 (bottom).
\label{spectrafig}}
\end{figure}

As shown in the left panel of Figure \ref{abshistfig}, absorption is common at
all luminosities, and over 50\% (30 of 53) of the low-redshift Type 1 AGN
observed using {\it FUSE} show detectable {\sc O~vi} absorption,
comparable to those Seyferts that show
longer-wavelength UV \citep{Crenshaw99}
or X-ray \citep{Reynolds97, George98} absorption.
None show intrinsic $\rm H_2$ absorption.
We see three basic morphologies for {\sc O~vi} absorption lines:
(1) {\bf Single}: 13 of 30 objects exhibit single, narrow,
isolated {\sc O~vi} absorption lines, as
illustrated by the spectrum of Ton~S180 \citep{Turner01}.
PG0804+761, shown in the top panel of Figure \ref{spectrafig}, is another
example.
(2) {\bf Blend}: multiple {\sc O~vi} absorption components that are blended
together.
10 of 30 objects fall in this class, and the spectrum of Mrk~279
is typical \citep{Scott04}.
The middle panel of Figure \ref{spectrafig} shows Mrk~478 as another example.
(3) {\bf Smooth}: The 7 objects here are an extreme expression of the
``blend" class,
where the {\sc O~vi} absorption is so broad and blended that individual
{\sc O~vi} components cannot be identified.  NGC~4151 typifies
this class \citep{Kriss92, Kriss95, Kriss01}.
The mini-BAL QSO PG1411+442 is another example of this class, shown in the
bottom panel of Figure \ref{spectrafig}.

Individual {\sc O~vi} absorption components in our spectra have
FWHM of 50--750 $\rm km~s^{-1}$, with most objects having
FWHM $< 100~\rm km~s^{-1}$.
The multiple components that are typically present are almost always
blue shifted, and they span a velocity range of 200--4000 $\rm km~s^{-1}$;
half the objects span a range of $< 1000~\rm km~s^{-1}$.
As shown in the right panel of Figure \ref{abshistfig}, the maximum outflow
velocities show a tendency to increase with source luminosity, perhaps
indicating that radiative acceleration plays some role in the dynamics.
Note also that there is a population of low-velocity absorbers present
at all luminosities.

\section{Mass Outflows from AGN}

The absorption lines we see in our spectra of AGN are associated with
mass outflows from the active nucleus.
These outflows can profoundly affect the
evolution of the central engine \citep{BB99}, the host galaxy
and its interstellar medium \citep{SR98, WL03}
and also the surrounding intergalactic medium (IGM) \citep{Cavaliere02,
Granato04, SO04}.  Winds from the high
metal-abundance nuclear regions may be a significant source for
enriching the IGM \citep{Adelberger03}.
Absorption by the outflow can also collimate the ionizing radiation
\citep{Kriss97} and thereby influence the ionization
structure of the host galaxy and the surrounding IGM.

The outflowing gas in AGN is sometimes visible as extended, bi-conical emission
at visible or X-ray wavelengths (e.g., NGC 4151, \citealt{Evans93},
\citealt{Hutchings98}; NGC 1068, \citealt{Ogle03}),
but it most frequently manifests itself as blue-shifted
absorption features in their UV and X-ray spectra.
About half of all low-redshift AGN show X-ray absorption
by highly ionized gas \citep{Reynolds97, George98}, and a similar
fraction show associated UV absorption in ionized species such
as {\sc C~iv} \citep{Crenshaw99} and {\sc O~vi} \citep{Kriss01}.
In more luminous quasars, the fraction of objects in the Sloan Digital Sky
Survey that shows broad {\sc C~iv} absorption troughs rises steeply as the
troughs become narrower \citep{Tolea02,Reichard03}, comprising over 30\% of
the quasar population at widths narrower than 1000 $\rm km~s^{-1}$.
For AGN that have been observed in both the X-ray and the UV,
there is a one-to-one correspondence between
objects showing X-ray and UV absorption, implying that the phenomena are
related in some way (Crenshaw et al. 1999).
The high frequency of occurrence of UV and X-ray absorption suggests that the
absorbing gas has a high covering fraction, and that it is present in all AGN.
The gas has a total mass exceeding $\sim 10^3~\rm M_\odot$ (greater than the
broad-line region, or BLR), and is outflowing at a rate
$>0.1~\rm M_\odot~yr^{-1}$
($10 \times$ the accretion rate in some objects) \citep{Reynolds97}.
In an effort to understand AGN outflows better, {\it FUSE} guest investigators have
been successful in coordinating several campaigns on bright AGN with
simultaneous {\it HST} and X-ray observations of NGC~3783, NGC~5548, NGC~4051,
NGC~4151, Ark~564, Mrk~279, and NGC~7469.

Some of the key results from these campaigns are:

\vspace{-6pt}
\begin{itemize}

\smallskip
\item 
Doublet ratios for the {\sc O~vi} absorption lines show that the absorbers can
be optically thick,
but they are not black at line center. Thus column densities are frequently
underestimated, sometimes by as much as an order of magnitude
\citep{Arav02, Arav03}.

\smallskip
\vspace{-6pt}
\item 
Variations in the absorption line strength can reflect either an ionization
response, or bulk motion
\citep{Crenshaw99, Gabel03b}.

\smallskip
\vspace{-6pt}
\item 
In the extensive recent {\it Chandra/FUSE/HST} campaign on NGC 3783
\citep{Kaspi02, Gabel03a}, the kinematics of the X-ray absorbing gas
provide a good match to the UV-absorbing gas.
The X-ray absorbing gas itself contains material spanning a large range
of ionization parameters \citep{Lee02, Sako03, Netzer03},
and it is likely that this broad range of physical
conditions can also include the UV-absorbing ions. This is a natural prediction
of the thermally driven wind model of \citet{KK95, KK01}, and would also be
likely in disk-driven winds.

\smallskip
\vspace{-6pt}
\item 
In other cases the UV gas visible in {\it FUSE} spectra appears to be in a fairly low
ionization state, and there is no direct relation between the
X-ray absorption and the multiple kinematic components seen in the UV
(NGC~5548: \citealt{Brotherton02};
Mrk~509: \citealt{Kriss00, Yaqoob03};
NGC~7469: \citealt{Kriss03, Blustin03}).

\end{itemize}

The multiple kinematic components frequently seen in the UV absorption spectra
of AGN clearly show that the absorbing medium is complex, with separate
UV and X-ray dominant zones.
In some cases, the UV absorption component corresponding to the X-ray warm
absorber can be clearly identified, such as
NGC~3783 (\citealt{Kaspi02, Gabel03a, Netzer03}).
In others, however, {\it no} UV absorption component shows physical
conditions characteristic of those seen in the X-ray absorber
as in NGC~5548 (\citealt{Brotherton02}).
One potential geometry for this complex absorbing structure is high-density,
low-column UV-absorbing clouds embedded in a low-density,
high-ionization medium that dominates the X-ray absorption.

Disk-driven winds are a possible explanation for some cases of AGN outflows
\citep{KK94, Murray95, Elvis00, Proga00}.
By analogy to stellar winds, one would expect the terminal velocity of an
AGN outflow to reflect the gravity of its origin.
Disk-driven winds should therefore have velocities in the range of several
thousand $\rm km~s^{-1}$.
Objects with broad, smooth profiles might fall in this category.
The geometry proposed by \citet{Elvis00} suggests that these objects should
have only modest inclinations. However, two prime examples of Seyferts with
broad smooth absorption troughs, NGC 3516 \citep{Hutchings01} and
NGC 4151 \citep{Kriss01}, are likely the
highest inclination sources in our sample given their extended, bi-conical
narrow emission-line region morphologies \citep{Miyaji92, Evans93} and their
opaque Lyman limits \citep{Kriss97}.

The lower velocities we observe in objects like NGC 3783 and NGC 5548
are more compatible with thermally driven winds
from the obscuring torus \citep{KK95, KK01}.
In these thermally driven winds, photoionized evaporation in the presence of
a copious mass source (the torus) locks the ratio of ionizing intensity to
total gas pressure (the ionization parameter $\Xi$) at a critical value.
For AGN spectral energy distributions lacking a strong extreme ultraviolet
bump, such as the composite spectra of quasars assembled by \citet{Zheng97},
\citet{Laor97}, and \citet{Telfer02},
the ionization equilibrium curve exhibits an extensive vertical branch.
Thus, at the critical ionization parameter for evaporation, there is a broad
range of temperatures that can coexist in equilibrium at nearly constant
pressure.
For this reason, the flow is expected to be strongly inhomogeneous.
Outflow velocities are typical of the sound speed in the heated gas, or
several hundred $\rm km~s^{-1}$, comparable to the velocities seen in many AGN.

In summary, we find that {\sc O~vi} absorption is common in low-redshift ($z < 0.15$)
AGN. 30 of 53 Type 1 AGN with $z < 0.15$ observed using {\it FUSE} show
multiple, blended {\sc O~vi} absorption lines with typical widths
of $\sim 100~\rm km~s^{-1}$ that are blueshifted over a velocity range of $\sim$
1000 $\rm km~s^{-1}$.
Those galaxies in our sample with existing X-ray or longer wavelength UV
observations also show {\sc C~iv} absorption and evidence of a soft X-ray
warm absorber.
In some cases, a UV absorption component has physical properties
similar to the X-ray absorbing gas, but in others there is no clear
physical correspondence between the UV and X-ray absorbing components.

\acknowledgements
I thank all the members of the {\it FUSE} AGN Working Group for their contributions
to this research, especially Jennifer Scott and Zhaohui Shang.
This work is based on data obtained for the Guaranteed Time Team by the
NASA-CNES-CSA FUSE mission operated by the Johns Hopkins University. Financial
support to U. S. participants has been provided by NASA contract NAS5-32985.


\begin{thebibliography}{}

\bibitem[Adelberger et al.(2003)]{Adelberger03}
Adelberger, K. L., et al. 2003, \apj, 584, 45

\bibitem[Arav, Korista, \& de Kool(2002)Arav et al.]{Arav02}
Arav, N., Korista, K. T., \& de Kool, M. 2002, \apj, 566, 699

\bibitem[Arav et al.(2003)]{Arav03}
Arav, N., et al. 2003, \apj, 590, 174

\bibitem[Blandford \& Begelman(1999)]{BB99}
Blandford, R, D., \& Begelman, M. C. 1999, \mnras, 303, L1

\bibitem[Blustin et al.(2003)]{Blustin03}
Blustin, A., et al. 2003, \aa, 403, 481

\bibitem[Brotherton et al.(2002)]{Brotherton02}
Brotherton, M., et al. 2002, \apj, 565, 800

\bibitem[Cavaliere, Lapi, \& Menci(2002)Cavaliere et al.]{Cavaliere02}
Cavaliere, A., Lapi, A., Menci, N., 2002, \apj, 581, L1

\bibitem[Crenshaw et al.(1999)]{Crenshaw99}
Crenshaw, D. M., et al. 1999, \apj, 516, 750


\bibitem[Elvis(2000)]{Elvis00}
Elvis, M.\ 2000, \apj, 545, 63


\bibitem[Evans et al.(1993)]{Evans93}
Evans, I., et al. 1993, \apj, 417, 82

\bibitem[Gabel et al.(2003a)]{Gabel03a}
Gabel, J., et al. 2003a, \apj, 583, 178

\bibitem[Gabel et al.(2003b)]{Gabel03b}
Gabel, J., et al. 2003b, \apj, 595, 120

\bibitem[George et al.(1998)]{George98}
George, I. M., et al. 1998, \apjs, 114, 73

\bibitem[Granato et al.(2004)]{Granato04}
Granato, G. L., et al. 2004, \apj, 600, 580

\bibitem[Hubeny et al.(2000)]{Hubeny00}
Hubeny, I., Agol, E., Blaes, O., \& Krolik, J. H. 2000 \apj, 533, 710

\bibitem[Hutchings et al.(1998)]{Hutchings98}
Hutchings, J., et al. 1998, \apj, 492, L115

\bibitem[Hutchings et al.(2001)]{Hutchings01}
Hutchings, J., et al. 2001, \apj, 559, 173

\bibitem[Kaspi et al.(2002)]{Kaspi02}
Kaspi, S., et al. 2002, \apj, 574, 643

\bibitem[K\"onigl \& Kartje(1994)]{KK94}
K\"onigl, A., \& Kartje, J.~F.\ 1994, \apj, 434, 446




\bibitem[Kriss(2001)]{Kriss01}
Kriss, G. A. 2001, in Mass Outflows in Active Galactic Nuclei: New
Perspectives, ed. D. M. Crenshaw, S. B. Kraemer, \& I. M. George,
A. S. P. Conference Series, 255, 69 (ASP: San Francisco)

\bibitem[Kriss et al.(1992)]{Kriss92}
Kriss G. A., et al. 1992, \apj, 392, 485

\bibitem[Kriss et al.(1995)]{Kriss95}
Kriss G. A., et al. 1995, \apj, 454, L7



\bibitem[Kriss et al.(1997)]{Kriss97}
Kriss G.~A., et al. 1997, in proceedings IAU Colloquium 159,
ed. B. M. Peterson, F.-Z. Cheng, \& A. S. Wilson,
A. S. P. Conference Series, 113, 453 (ASP: San Francsico)


\bibitem[Kriss et al.(2000)]{Kriss00}
Kriss, G. A., et al. 2000, \apj, 538, L17

\bibitem[Kriss et al.(2003)]{Kriss03}
Kriss, G. A., et al. 2003, \aa, 403, 473

\bibitem[Krolik \& Kriss(1995)]{KK95}
Krolik, J. H., \& Kriss, G. A. 1995, \apj, 447, 512

\bibitem[Krolik \& Kriss(2001)]{KK01}
Krolik, J. H., \& Kriss, G. A. 2001, \apj, 561, 684

\bibitem[Laor et al.(1997)]{Laor97}
Laor, A., et al. 1997, \apj, 477, 93

\bibitem[Lee et al.(2002)]{Lee02}
Lee, J. C., et al. 2002, in X-ray Spectroscopy of AGN with Chandra
and XMM-Newton, ed. T. B\"oller (http://www.xray.mpe.mpg.de/$\sim$bol/agnspec/programm.html)





\bibitem[Murray et al.(1995)]{Murray95}
Murray, N., et al. 1995, \apj, 451, 498

\bibitem[Miyaji, Wilson, \& Perez-Fournon(1992)Miyaji et al.]{Miyaji92}
Miyaji, T., Wilson, A., \& Perez-Fournon, I. 1992, \apj, 385, 137

\bibitem[Netzer et al.(2003)]{Netzer03}
Netzer, H., et al.\ 2003, \apj, 599, 933

\bibitem[Ogle et al.(2003)]{Ogle03}
Ogle, P., et al. 2003, \aa, 402, 849

\bibitem[Proga(2000)]{Proga00}
Proga, D. 2000, \apj, 538, 684

\bibitem[Reichard et al.(2003)]{Reichard03}
Reichard, T. A., et al. 2003, \aj, 125, 1711

\bibitem[Reynolds(1997)]{Reynolds97}
Reynolds, C. S. 1997, \mnras, 286, 513

\bibitem[Sako et al.(2003)]{Sako03}
Sako, M., et al. 2003, \apj, 596, 114

\bibitem[Scannapieco \& Oh(2004)]{SO04}
Scannapieco, E. \& Oh, S. P. 2004, \apj, 608, 62

\bibitem[Scott et al.(2004)]{Scott04}
Scott, J. E., et al. 2004, \apj, 615, 135

\bibitem[Shang et al.(2004)]{Shang04}
Shang, Z., et al. 2004, \apj\ in press (astro-ph/0409697).

\bibitem[Silk \& Rees(1998)]{SR98}
Silk, J., \& Rees, M. J. 1998, \aa, 331, L1S.

\bibitem[Telfer et al.(2002)]{Telfer02}
Telfer, R. C., et al. 2002, \apj, 565, 773

\bibitem[Tolea, Krolik, \& Tsvetanov(2002)Tolea et al.]{Tolea02}
Tolea, A., Krolik, J. H., \& Tsvetanov, Z. 2002, \apj, 578, L31

\bibitem[Turner et al.(2001)]{Turner01}
Turner, T. J., et al. 2001, \apj, 548, L13

\bibitem[Wyithe \& Loeb(2003)]{WL03}
Wyithe, J.~S.~B.~\& Loeb, A.\ 2003, \apj, 595, 614

\bibitem[Yaqoob et al.(2003)]{Yaqoob03}
Yaqoob, T., et al. 2003, \apj, 582, 105

\bibitem[Zheng et al.(1997)]{Zheng97}
Zheng, W., et al. 1997, \apj, 475, 469

\end{thebibliography}
\end{document}